\documentclass{amsart}

\usepackage{amsfonts}
\usepackage[foot]{amsaddr}
\usepackage{amsmath}
\usepackage{array}
\usepackage{subcaption}
\usepackage{stfloats}
\usepackage{url}
\usepackage{tikz}
\usepackage{lipsum,adjustbox}
\usepackage{verbatim}
\usepackage{algorithm}
\usepackage[noend]{algpseudocode}
\usepackage{pgfplots}
\pgfplotsset{compat=newest}
\usepackage{etoolbox}

\algnewcommand\algorithmicforeach{\textbf{for each}}
\algdef{S}[FOR]{ForEach}[1]{\algorithmicforeach\ #1\ \algorithmicdo}

\calclayout
\patchcmd{\subsection}{\bfseries}{\itshape}{}{}
\patchcmd{\subsection}{-.5em}{.5em}{}{}

\makeatletter
\renewcommand{\email}[2][]{%
  \ifx\emails\@empty\relax\else{\g@addto@macro\emails{,\space}}\fi%
  \@ifnotempty{#1}{\g@addto@macro\emails{\textrm{(#1)}\space}}%
  \g@addto@macro\emails{#2}%
}
\makeatother

\title[Batched QR and SVD Algorithms]{Batched QR and SVD Algorithms on GPUs with Applications in Hierarchical Matrix Compression}%

\author{Wajih Halim Boukaram$^1$}
\author{George Turkiyyah$^2$}
\author{Hatem Ltaief$^1$}
\author{David E. Keyes$^1$}

\email{wajihhalim.boukaram@kaust.edu.sa, gt02@aub.edu.lb, hatem.ltaief@kaust.edu.sa, \\
david.keyes@kaust.edu.sa}

\address{$^1$Extreme Computing Research Center (ECRC), King Abdullah University of Science and Technology (KAUST), Thuwal 23955, Saudi Arabia.}
\address{$^2$Department of Computer Science, American University of Beirut (AUB), Beirut, Lebanon.}

\begin{document}

\begin{abstract}
We present high performance implementations of the QR and the singular value decomposition of a batch of small matrices hosted on the GPU with applications in the compression of hierarchical matrices. The one-sided Jacobi algorithm is used for its simplicity and inherent parallelism as a building block for the SVD of low rank blocks using randomized methods. We implement multiple kernels based on the level of the GPU memory hierarchy in which the matrices can reside and show substantial speedups against streamed cuSOLVER SVDs. The resulting batched routine is a key component of hierarchical matrix compression, opening up opportunities to perform H-matrix arithmetic efficiently on GPUs.
\end{abstract}

\maketitle

\section{Introduction}
\label{sec:intro}
The singular value decomposition (SVD) is a factorization of a general $m \times n$ matrix $A$ of the form
\begin{equation*}
A = U \Sigma V^*.
\end{equation*}
$U$ is an $m \times m$ orthonormal matrix whose columns $U_i$ are called the left singular vectors. $\Sigma$ is an $m \times n$ diagonal matrix whose diagonal entries $\sigma_i$ are called the singular values and are sorted in decreasing order. $V$ is an $n \times n$ orthonormal matrix whose columns $V_i$ are called the right singular vectors. When $m > n$, we can compute a reduced form $A = \hat{U} \hat{\Sigma} V^*$ where $\hat{U}$ is an $m \times n$ matrix and $\hat{\Sigma}$ is an $n \times n$ diagonal matrix. One can easily obtain the full form from the reduced one by extending $\hat{U}$ with $(m - n)$ orthogonal vectors and $\hat{\Sigma}$ with an $(m - n)$ zero block row. Without any loss of generality, we will focus on the reduced SVD of real matrices in our discussions.

The SVD of a matrix is a crucial component in many applications in signal processing and statistics as well as matrix compression, where truncating the $(n - k)$ singular values that are smaller than some threshold gives us a rank-$k$ approximation $\tilde{A}$ of the matrix $A$. This matrix is the unique minimizer of the function $f_k(B) = || A - B ||_F$. In the context of hierarchical matrix operations, effective compression relies on the ability to perform the computation of large batches of independent SVDs of small matrices of low numerical rank. Randomized methods \cite{halko2011finding} are well suited for computing a truncated SVD of these types of matrices and are built on three computational kernels: the QR factorization, matrix-matrix multiplications and SVDs of smaller $k \times k$ matrices. Motivated by this task, we discuss the implementation of high performance batched QR and SVD kernels on the GPU, focusing on the more challenging SVD tasks.  

The remainder of this paper is organized as follows. Section \ref{sec:background} presents different algorithms used to compute the QR factorization and the SVD as well as some considerations when optimizing for GPUs. Section \ref{sec:batch_qr} discusses the batched QR factorization and compares its performance with existing libraries. Sections \ref{sec:registers}, \ref{sec:shared} and \ref{sec:block_global} discuss the various implementations of the SVD based on the level of the memory hierarchy in which the matrices can reside. Specifically, Section \ref{sec:registers} describes the implementation for very small matrix sizes that can fit in registers, Section \ref{sec:shared} describes the implementation for matrices that can reside in shared memory, and Section \ref{sec:block_global} describes the block Jacobi implementation for larger matrix sizes that must reside in global memory. Section \ref{sec:randomized} details the implementation of the batched randomized SVD routine. We then discuss some details of the application to hierarchical matrix compression in Section \ref{sec:application}. We conclude and discuss future work in Section \ref{sec:conclusion}.

\section{Background}
\label{sec:background}

In this section we give a review of the most common algorithms used to compute the QR factorization and the SVD of a matrix as well as discuss some considerations when optimizing on the GPU.

\subsection{QR Factorization}
The QR factorization decomposes an $m \times n$ matrix $A$ into the product of an orthogonal $m \times m$ matrix $Q$ and an upper triangular $m \times n$ matrix $R$ \cite{golub2013matrix}. We can also compute a reduced form of the decomposition where Q is an $m \times n$ matrix and R is $n \times n$ upper triangular. The most common QR algorithm is based on transforming $A$ into an upper triangular matrix using a series of orthogonal transformations generated using Householder reflectors. Other algorithms such as the Gram-Schmidt or Modified Gram-Schmidt can produce the QR factorization by orthogonalizing a column with all previous columns; however, these methods are less stable than the Householder orthogonalization and the orthogonality of the resulting $Q$ factor suffers with the condition number of the matrix. Another method is based on Givens rotations, where entries in the subdiagonal part of the matrix are zeroed out to form the triangular factor and the rotations are accumulated to form the orthogonal factor. This method is very stable and has more parallelism than the Householder method; however it is more expensive, doing about 50\% more work, and it is more challenging to extract the parallelism efficiently on the GPU. For our implementation, we rely on the Householder method due to its numerical stability and simplicity. The method is described in pseudo-code in Algorithm \ref{alg:qr}.
\begin{algorithm} [t]
\caption{Householder QR}
\label{alg:qr}
	\begin{algorithmic}[1]
		\Procedure{QR}{$A, Q, R$}
		\State $[Q, R] = [I, A]$
		\For{$i = 1 \rightarrow A.n$}
			\State $v = \text{house}(R(i))$
			\State $R = (I - 2vv^T) R$  \label{alg:qr:trailing_update}
			\State $Q = Q (I - 2vv^T)$
		\EndFor
		\EndProcedure
	\end{algorithmic}
\end{algorithm}	

\subsection{SVD Algorithms}
Most implementations of the SVD are based on the two-phase approach popularized by
Trefethen et al.~\cite{trefethen1997numerical}, where the matrix $A$ first undergoes bidiagonalization of
the form $A = Q_U B Q_V^T$ where $Q_U$ and $Q_V$ are orthonormal matrices and $B$ is a 
bidiagonal matrix. The matrix $B$ is then diagonalized using some variant of the QR 
algorithm, the divide and conquer method or a combination of both to produce a 
decomposition $B = U_B \Sigma V_B^T$. The complete SVD is then determined as $A = (Q_U U_B) 
\Sigma (Q_V V_B)^T$ during the backward transformation. These methods require significant algorithmic and programming effort 
to become robust and efficient while still suffering from a loss of relative accuracy 
\cite{demmel1992jacobi}. 

An alternative is the one-sided Jacobi method where all $n(n-1)/2$ pairs of columns are repeatedly
orthogonalized in sweeps using plane rotations until all columns are mutually orthogonal. When the process
converges (i.e., all columns are mutually orthogonal up to machine precision), the left singular vectors are the normalized columns 
of the modified matrix with the singular values as the norms of those columns. The right singular vectors can
be computed either by accumulating the rotations or by solving a system of equations. Our application does
not need the right vectors, so we omit the details of computing them. Algorithm \ref{alg:jacobi} describes
the one-sided Jacobi method. Since each pair of columns can be orthogonalized independently, the method is
also easily parallelized. The simplicity and inherent parallelism of the method make it an attractive first
choice for an implementation on the GPU.

\begin{algorithm} [b]
\caption{One-sided Jacobi SVD}
\label{alg:jacobi}
	\begin{algorithmic}[1]
		\While {not converged}
			\ForEach {pair of columns $A_{ij}=[A_i, A_j]$}
				\State $G = A_{ij}^T A_{ij}$ \label{alg:jacobi:gram}
				\State $R = rot(G)$
				\State $A_{ij} = A_{ij} R$	 \label{alg:jacobi:rot}
			\EndFor
		\EndWhile
	\end{algorithmic}
\end{algorithm}	

\subsection{GPU Optimization Considerations}
GPU kernels are launched by specifying a grid configuration which lets us organize threads into 
blocks and blocks into a grid. Launching a GPU kernel causes a short stall (as much as 10 microseconds) 
as the kernel is prepared for execution. This kernel launch overhead prevents kernels that complete their 
work faster than the overhead from executing in parallel, essentially serializing them. 
To overcome this limitation when processing small workloads, the work is batched into a 
single kernel call when possible \cite{batchqr_haidar, batch_haidar}.  
All operations can then be executed in parallel without incurring the kernel launch overhead, 
with the grid configuration used to determine thread work assignment.

A warp is a group of threads (32 threads in current generation GPUs, such as the NVIDIA 
K40) within a block that executes a single instruction in lockstep, without requiring any explicit
synchronization. The occupancy of a kernel tells us the ratio of active warps to the 
maximum number of warps that a multiprocessor can host. This metric is dependent on the 
amount of resources that a kernel uses, such as register and shared memory usage and 
kernel launch configuration, as well as the compute capability of the card 
(\cite{wilt2013cuda} for more details). While not a requirement for good 
performance \cite{volkov2010better}, it is generally a good idea to aim for high 
occupancy.

\begin{figure}[t]
\centering
\includegraphics[width=0.45\textwidth]{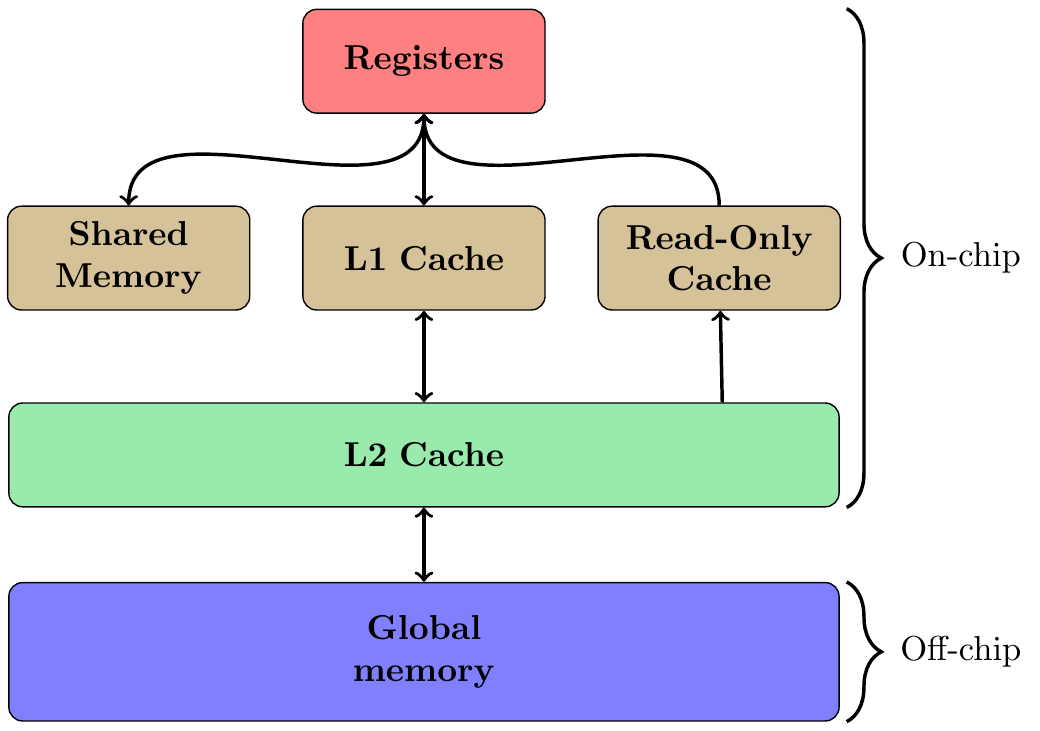}
\caption{The memory hierarchy of a modern GPU.}
\label{fig:memory_hierarchy}
\end{figure}

Memory on the GPU is organized into a hierarchy of memory spaces as shown in Figure 
\ref{fig:memory_hierarchy}. At the bottom, we have global memory which is accessible by all threads 
and is the most plentiful but the slowest memory. The next space of interest is the shared memory 
which is accessible only by threads within the same block and is configurable with the L1 cache to 
be at most 48KB per thread block on current generation GPUs. Shared memory is very fast and acts as a programmer 
controllable cache. Finally, we have the registers which are local to the threads. Registers are the  
fastest of all memory, but the total number of registers usable by a thread without performance 
implications is limited. If a kernel needs more registers than the limit, then registers are 
spilled to ``local" memory, which is in the slow but cached global memory. 
Making good use of the faster memories and avoiding excessive accesses to the slower ones is key to 
good performance on the GPU. As such, it is common to use blocking techniques in many algorithms, 
where a block of data is brought in from global memory and processed in one of the faster memories.

\subsection{Related Work}
Batched GPU routines for LU, Cholesky and QR factorizations have been developed in \cite{batchqr_haidar, batch_haidar, charara_batch_tdla} using a block recursive approach which increases data reuse and leads to very good performance for relatively large matrix sizes. GPU routines optimized for computing the QR decomposition of very tall and skinny matrices are presented in \cite{caqr_anderson} where they develop an efficient transpose matrix-vector computation that is employed with some minor changes in this work. GPU-CPU hybrid algorithms for batched SVD using Jacobi and bidiagonalization methods are introduced in \cite{kotas_svd} where pair generation for the Jacobi method and the solver phase of the bidiagonalization are handled on the CPU. The work in \cite{Kang2015} employs the power method to construct a rank 1 approximation for 2D filters in convolutional neural networks. Routines to handle the SVD of many matrices on GPUs is presented in \cite{badolato_2015} where each thread within a warp computes the SVD of a single matrix.


\section{Batched QR Decomposition}
\label{sec:batch_qr}
In this section, we discuss implementation details of our batched QR kernel and compare it with other implementations from the MAGMA 2.2 \cite{tnld10} and CUBLAS 8 \cite{nvidia-cublas} libraries. 

\subsection{Implementation}
\begin{figure} 
	\centering
    \includegraphics[width=0.6\linewidth]{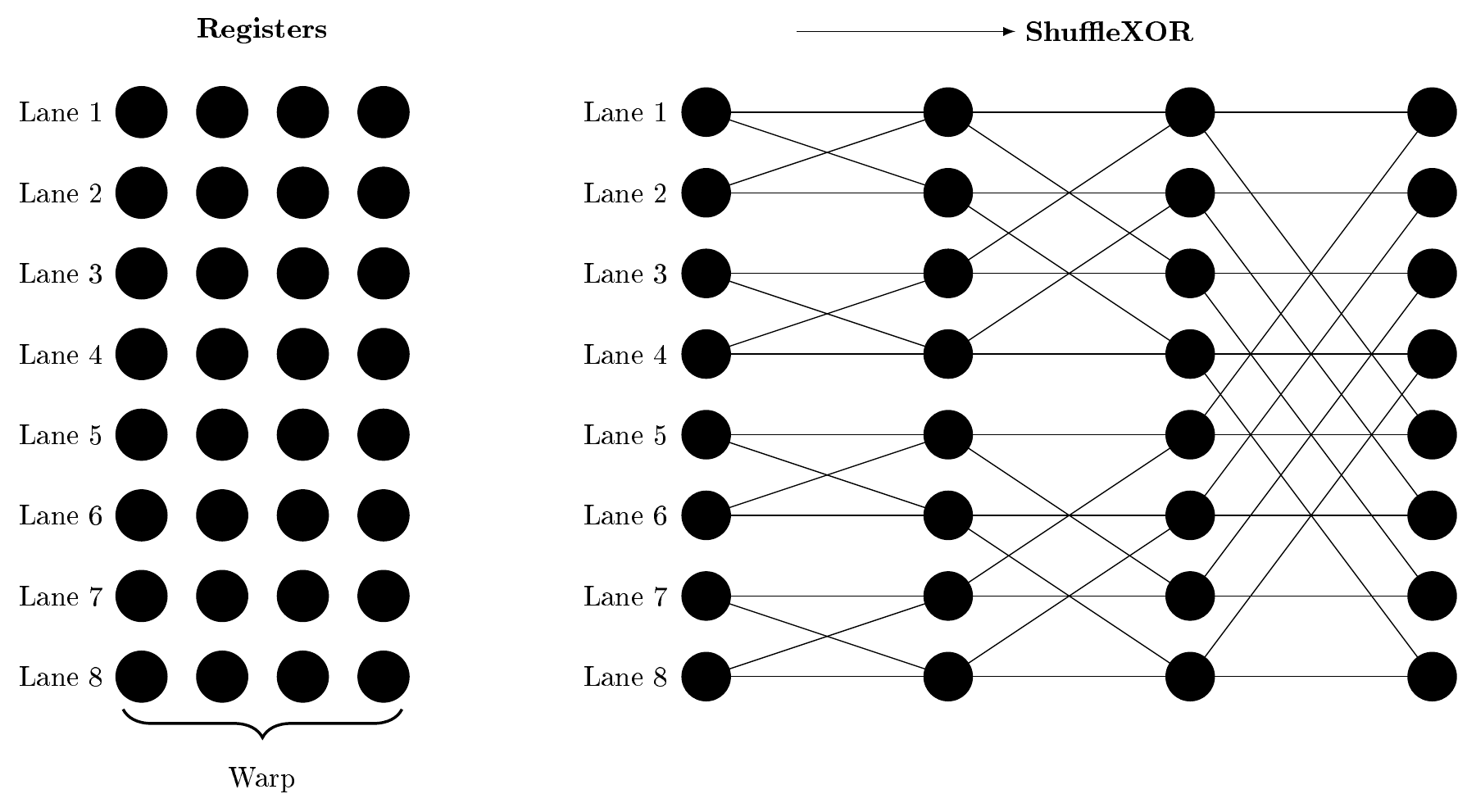}
    \caption{Left: matrix rows allocated to thread registers in a warp. Right: parallel warp reduction using shuffles within registers.}
    \label{fig:register_storage_reduction}
\end{figure}

\begin{figure}[t!]
\centering
\includegraphics[width=0.65\textwidth]{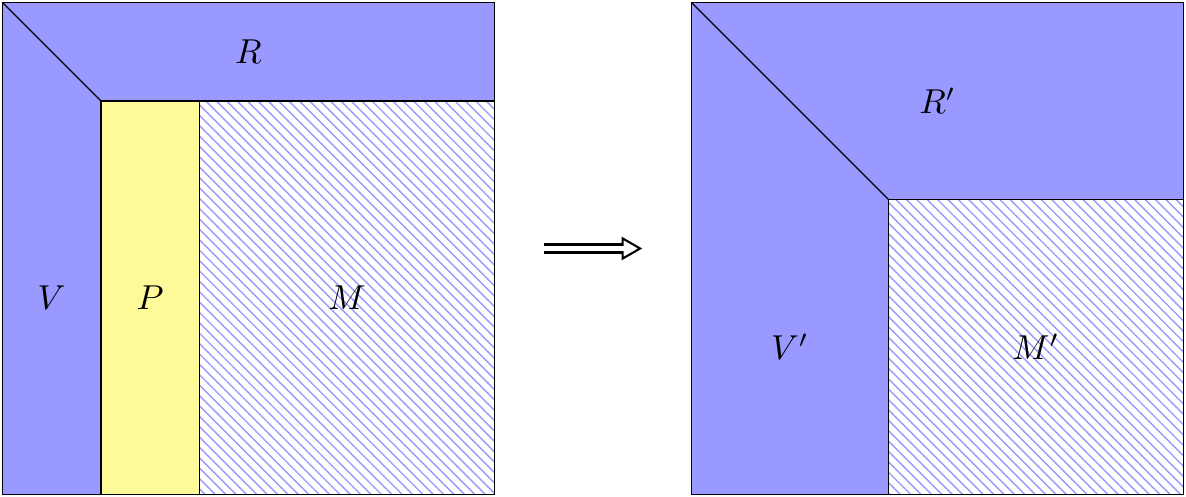}
\caption{One step of the QR factorization where a panel P is factored to produce a triangular factor R and reflectors V which are used to update the trailing sub-matrix M.}
\label{fig:qr_fig}
\end{figure}

One benefit of the Householder algorithm is that the application of reflectors to the trailing matrix (line \ref{alg:qr:trailing_update} of the algorithm) can be blocked together and expressed as a matrix-matrix multiplication (Level 3 BLAS) instead of multiple matrix-vector multiplications (Level 2 BLAS). The increased arithmetic intensity typically allows performance to improve when the trailing matrix is large. However, for small matrix blocks, the overhead of generating the blocked reflectors from their vector form as well as the lower performance of the matrix-matrix multiplication for small matrices hinder performance. We can obtain better performance by applying multiple reflectors in their vector form and performing the transpose matrix-vector multiplication efficiently within a thread block \cite{caqr_anderson}. First, we perform the regular factorization on a column block $P$ (called a panel). The entire panel is stored in registers, with each thread storing one row of the panel, and the transpose matrix-vector product is computed using a series of reductions using shared memory and warp shuffles \cite{warp_shfl} which allow threads within a warp to read each other's registers. Figure \ref{fig:register_storage_reduction} shows the data layout for a theoretical warp of size 8 with 4 columns in registers and a warp reduction using shuffles. Once we factor the panel, we can apply the reflectors to the trailing sub-matrix in a separate kernel that is optimized for performing the core matrix-vector product in the update. In this second kernel, we load both the factored panel $P$ and a panel $M_i$ of the trailing sub-matrix $M$ to registers and apply the reflectors one at a time, updating the trailing panel in registers. Let us take an example of a $32 \times 8$ trailing panel $M_i$. For each reflector, we compute the matrix-vector product $M_i^Tv$ by flattening the $32 \times 8$ product into a reduction of a 256 vector in shared memory that has been padded to avoid bank conflicts. The reduction can then be serialized until it reaches a size of 32, where a partial reduction to a vector of size 8 can take place in 2 steps. This final vector is the product $M_i^Tv$ which can then be quickly applied to the registers storing $M_i$. This process is repeated for each trailing panel within the same kernel to maximize the use of the reflectors which have been stored in registers. Figure \ref{fig:qr_fig} shows one step of a panel factorization and the application of its reflectors to the trailing submatrix. Since threads are limited to 1024 per block on current architectures, we use the approach developed in \cite{journals/concurrency/KurzakLDB10} to factorize larger matrices. We first factorize panels up to the thread block limit in a single kernel call. The panels below the first are then factorized by first loading the triangular factor into shared memory and then proceeding with the panel factorization as before, taking the triangular portion into consideration when computing reflectors and updates. To keep occupancy up for the small matrices on devices where the resident block limit could be reached before the thread limit, we assign multiple operations to a single thread block. For a batch of $N$ matrices of dimensions $m \times n$, kernels can be launched using $N/b$ thread blocks of size $m \times b$, where each thread block handles $b$ operations. 

\subsection{Performance}
\begin{figure}
	\begin{subfigure}[t]{0.45\textwidth}
	\centering
	\includegraphics{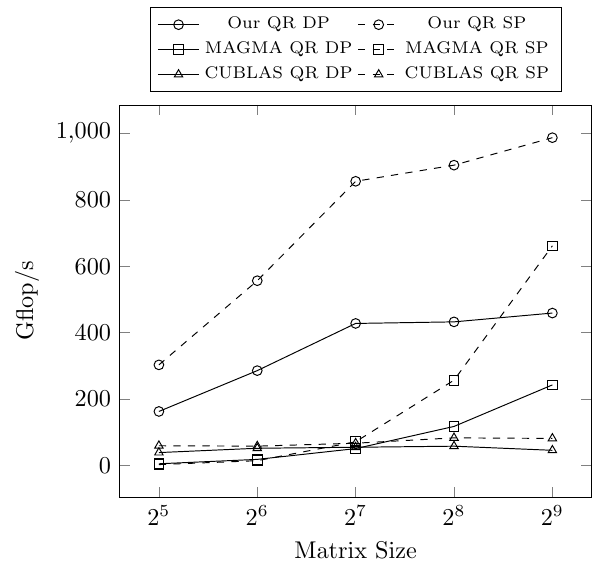}
	\caption{Batched QR kernel performance for square matrices.}
	\label{fig:batch_qr}
	\end{subfigure}
	\hfill	
	\begin{subfigure}[t]{0.45\textwidth}
	\centering
	\includegraphics{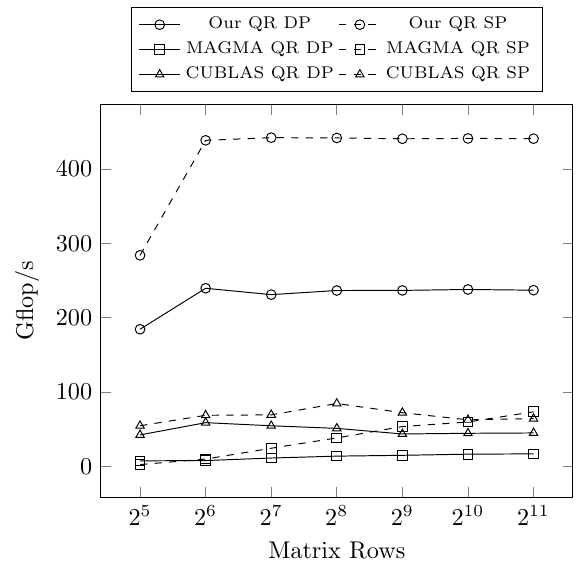}
	\caption{Batched QR kernel performance for rectangular matrices with a fixed column size of 32.}
	\label{fig:batch_qr_rect}
	\end{subfigure}
	\caption{Comparing batched QR kernels for 1000 matrices of varying size on a P100 GPU in single and double precision.}
\end{figure}

Figures \ref{fig:batch_qr} and \ref{fig:batch_qr_rect} show the performance of our batched QR for 1000 square and rectangular matrices with a panel width of $16$, tuned for the P100 GPU. We compare against the vendor implementation in CUBLAS as well as the high performance library MAGMA. We can see that our proposed version performs well for rectangular matrices with column size of 32 and starts losing ground against MAGMA for the larger square matrix sizes where the blocked algorithm starts to show its performance benefits. A nested implementation where our kernel can be used to factor relatively large panels in a blocked algorithm will likely show some additional performance improvements for the large square matrices, but we leave that as future work.

\section{Register Memory One-Sided Jacobi}
\label{sec:registers}
In this section we will discuss the first batched SVD kernel where the matrix data is hosted in registers and analyze the performance of the resulting kernel.

\subsection{Implementation}
In this implementation, to avoid repeated global memory accesses, we attempt to fit the matrix in register memory using the same layout as the panel in the QR factorization, i.e.\, one row per thread; however, the number of registers that a thread uses has an impact on occupancy which can potentially lead to lower performance. In addition, once the register count exceeds the limit set by the GPU's compute capability, the registers spill into ``local" memory which resides in cached slow global memory. Since we store an entire matrix row in the registers of one thread, we use the serial one-sided Jacobi algorithm to compute the SVD where column pairs are processed by the threads one at a time. The bulk of the work lies in the computation of the Gram matrix $G = A_{ij}^T A_{ij}$ (line \ref{alg:jacobi:gram} of Algorithm \ref{alg:jacobi}) and in the update of the columns (line \ref{alg:jacobi:rot}). Since the Gram matrix is symmetric, this boils down to three dot products which are executed as parallel reductions within the warp using warp shuffles. The computation of the $2 \times 2$ rotation matrix as well as the convergence test is performed redundantly in each thread. Finally, the column update is done in parallel by each thread on its own register data. As with the QR kernel, we keep occupancy up for the smaller matrix sizes by assigning multiple SVD operations to a single block of threads with each operation assigned to a warp to avoid unnecessary synchronizations. 

\subsection{Performance}
\label{subsec:reg_perf}
\begin{figure}
	
	\begin{subfigure}[t]{.45\linewidth}
	\centering
	\includegraphics{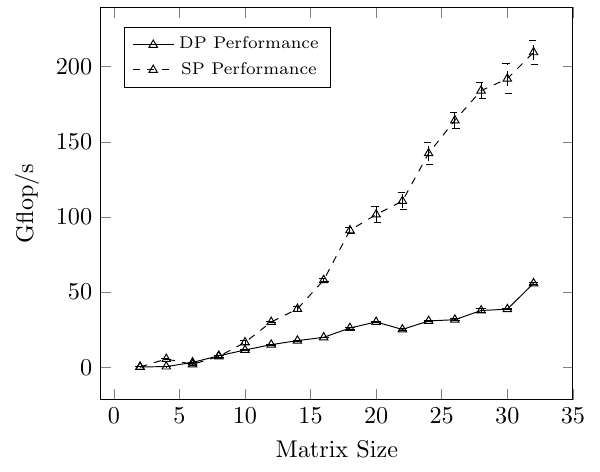}
	\caption{Kernel performance in GFLOP/s and achieved occupancy.}
	\label{fig:reg_svd_perf}
	\end{subfigure}
	\hfill
	\begin{subfigure}[t]{.45\linewidth}
	\centering
	\includegraphics{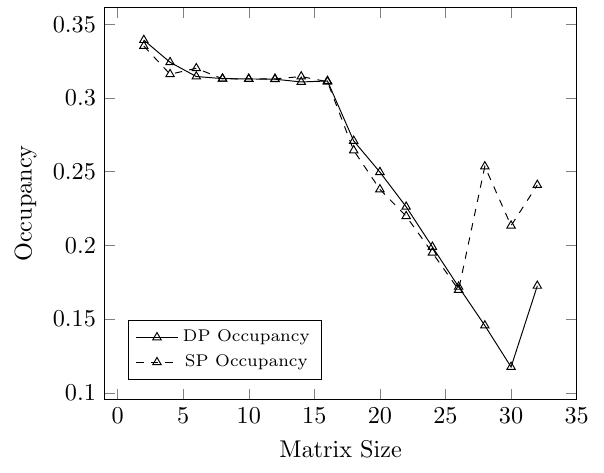}
	\caption{The effect of increasing the matrix size on the occupancy of the register kernel.}
	\label{fig:reg_svd_occupancy}
	
	\end{subfigure}
	\caption{Performance of the batched register memory SVD on a P100 GPU for 1000 matrices of varying size in single and double precision arithmetics.}
\end{figure}

We generate batches of 1000 test matrices with varying condition numbers using the \texttt{latms} LAPACK routine and calculate performance based on the total number of rotations needed for convergence. Figures \ref{fig:reg_svd_perf} and \ref{fig:reg_svd_occupancy} show the performance on a P100 GPU of the register-based batched SVD kernel and the effect increased register usage has on occupancy. Profiling the kernel, we see that the Gram matrix computation takes about 500 cycles, column rotations take about 240 cycles, and the redundantly computed convergence test and rotation matrices dominate at 1900 cycles. The fact that the redundant portion of the computation dominates means that it is preferable to assign as few threads as possible when processing column pairs. Due to the low occupancy for the larger matrix sizes and the register spills to local memory for matrices larger than 30, it is obvious that the register approach will not suffice for larger matrix sizes. This leads us to our next implementation based on the slower but more parallel-friendly shared memory.

\section{Shared Memory One-Sided Jacobi}
\label{sec:shared}
While the register based SVD performs well for very small matrix sizes, we need a kernel that can handle larger sizes and maintain reasonably high occupancy. This leads us to building a kernel based on shared memory, the next level of the GPU memory hierarchy. This section discusses the implementation details of this kernel and analyze its performance when compared with the register kernel.

\subsection{Implementation}
\begin{figure}[t]
	\centering
    \includegraphics[width=0.6\linewidth]{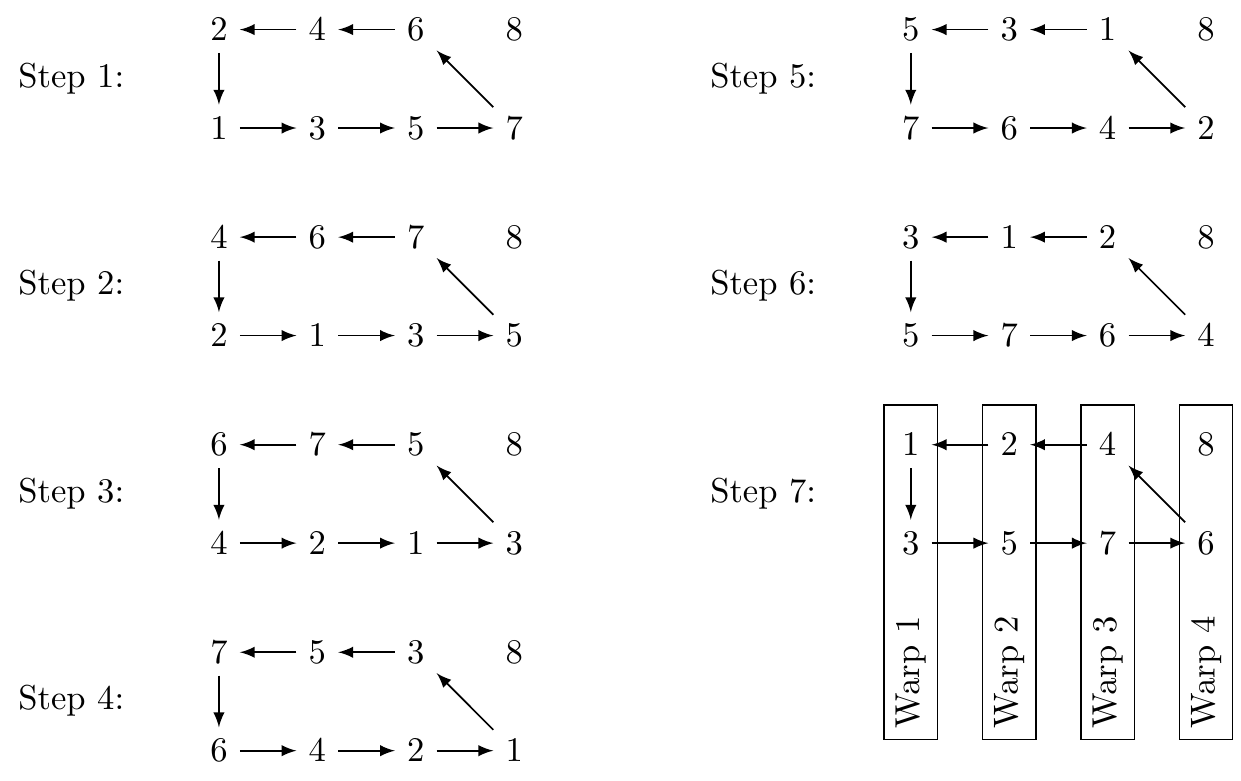}
    \caption{Distribution of column pairs to warps at each step of a sweep.}
    \label{fig:round_robin}
\end{figure}

\begin{figure} 
\centering
\begin{subfigure}[t] {0.3\linewidth}
	\centering
    \includegraphics[width=\linewidth]{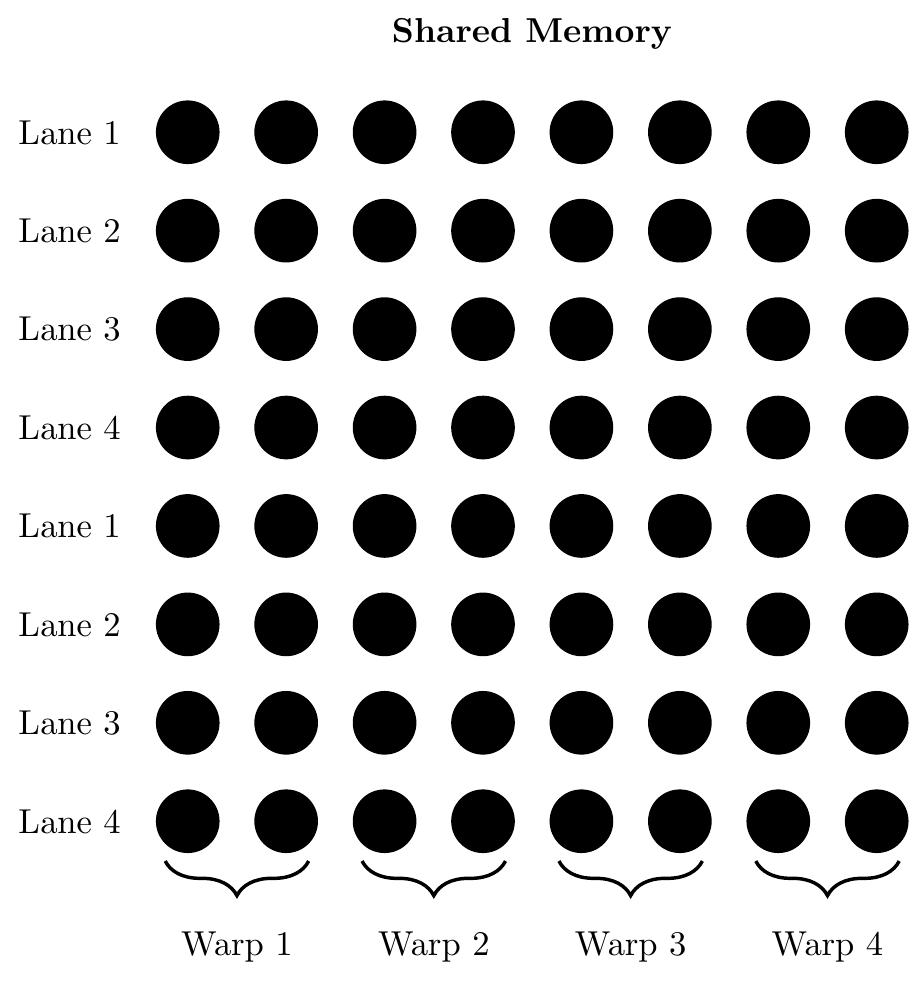}
    \caption{Matrix columns assigned in pairs to multiple warps and stored in shared memory.}
    \label{fig:shared_alloc}
\end{subfigure}
\hspace{2em}
\begin{subfigure}[t] {0.4\linewidth}
	\centering
    \includegraphics[width=\linewidth]{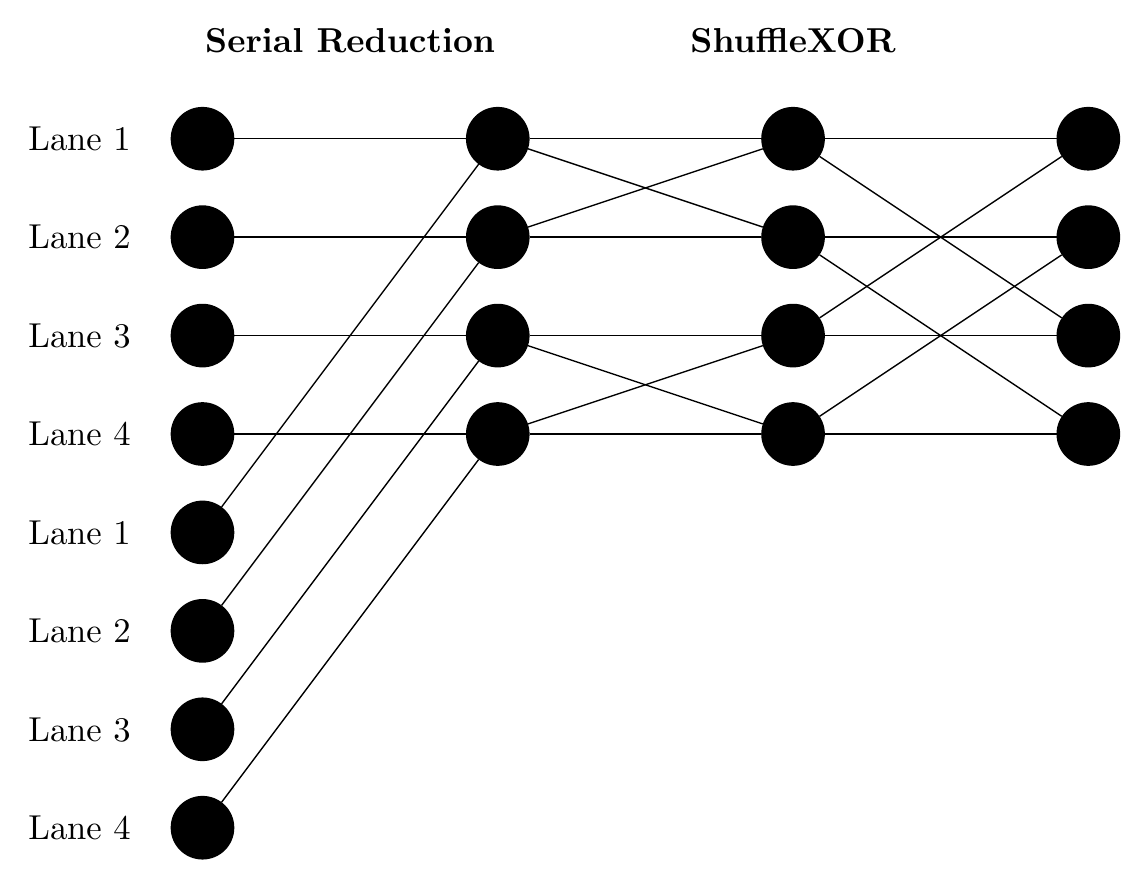}
    \caption{Parallel reduction of a column of data in shared memory using register shuffles after an initial serial reduction step.}
	\label{fig:shared_warp_reduction}
\end{subfigure}
\caption{Shared memory kernel implementation details.}
\end{figure}

In this version, the matrix is stored entirely in shared memory, which is limited to at most 48 KB per thread block on current generation GPUs. Using the same thread assignment as the register based kernel would lead to very poor occupancy due to the high shared memory consumption, where potentially only a few warps will be active in a multiprocessor. Instead, we exploit the inherent parallelism of the one-sided Jacobi to assign a warp to a pair of columns, i.e., there are $n/2$ warps processing an $m \times n$ matrix stored in shared memory. There are a total of $n(n-1)/2$ pairs of columns, so we must generate all pairings in $n-1$ steps, with each step processing $n/2$ pairs in parallel. There are many ways of generating these pairs, including round robin, odd-even, and ring ordering \cite{parosbj_zhou, ZHOU19971}. We implement the round robin ordering using shared memory to keep track of the column indexes of the pairs with the first warp in the block responsible for updating the index list after each step. Figure \ref{fig:round_robin} shows this ordering for a matrix with 8 columns. When the number of matrix rows exceeds the size of the warp, the thread-per-row assignment no longer allows us to use fast warp reductions, which would force us to use even more resources, as the reductions would now have to be done in shared memory. Instead, we assign multiple rows to a thread, serializing a portion of the reduction over those rows until warp reductions can be used. This follows our observation in Section \ref{subsec:reg_perf} to assign as few threads as possible to process column pairs, frees up valuable resources and increases the overall performance of the reduction. Row padding is used to keep the rows at multiples of the warp size, and column padding is used to keep the number of columns even. Kernels can then be launched using $32\times n/2$ threads to process each matrix. Figures \ref{fig:shared_alloc} and \ref{fig:shared_warp_reduction} show examples of the thread allocation and reductions for a $8 \times 8$ matrix using a theoretical warp size of 4.

\subsection{Performance}
\begin{figure}
	\begin{subfigure}[t]{0.45\linewidth}
	\centering
	\includegraphics{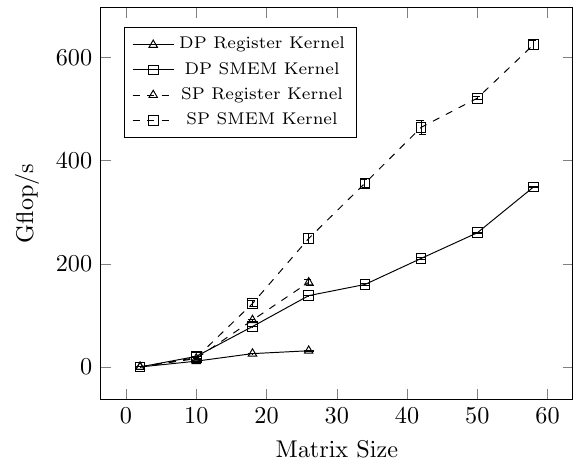}
	\caption{Shared memory kernel performance in GFLOPs/s compared to the register kernel.}
	\label{fig:shared_svd_perf}
	\end{subfigure}
	\hfill
	\begin{subfigure}[t]{0.45\linewidth}
	\centering
	\includegraphics{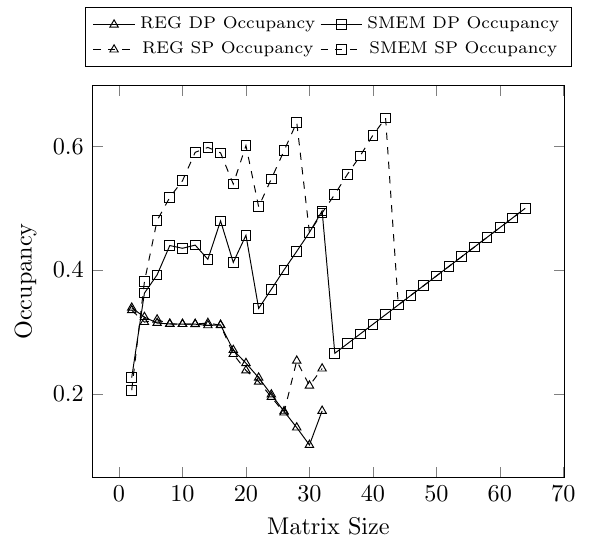}
	\caption{Comparison of the occupancy achieved by the register and shared memory kernels.}
	\label{fig:shared_svd_occupancy}
	\end{subfigure}
	\caption{Performance of the batched shared memory SVD on a P100 GPU for 1000 matrices of varying size in single and double precision arithmetics.}
\end{figure}

Figures \ref{fig:shared_svd_perf} and \ref{fig:shared_svd_occupancy} show the performance of the parallel shared SVD kernel compared to the serial register SVD kernel on a P100 GPU. We can see the improved growth in performance in the shared memory kernel due to the greater occupancy as well as the absence of any local memory transactions. Looking at the double precision occupancy, we notice two dips in occupancy at matrix sizes 22 and 32 as the number of resident blocks become limited by the registers/block limits of the device, dropping to 2 and then 1 resident blocks. Performance increases steadily from there as we increase the number of threads assigned to the operation until we reach a matrix size of $64 \times 64$ where we reach the block limit of 1024 threads. To handle larger sizes, we must use a blocked version of the algorithm or the randomized SVD as we see in Sections \ref{sec:block_global} and \ref{sec:randomized}, respectively. 

\section{Global Memory One-Sided Block Jacobi}
\label{sec:block_global}

When we can no longer store the entire matrix in shared memory, we have to operate on the matrix in the slower global memory. Instead of repeatedly reading and updating the columns one at a time, block algorithms that facilitate cache reuse have been developed \cite{bevcka1999block1, bevcka1999block2, bevcka2015new}. The main benefit of the block Jacobi algorithm is its high degree of parallelism; however, since we implement a batched routine for independent operations, we will use the serial block Jacobi algorithm for individual matrices and rely on the parallelism of the batch processing. The parallel version, where multiple blocks are processed simultaneously, can still be used when the batch size is very small, but we will focus on the serial version. In this section we will discuss the implementation details for two global memory block Jacobi algorithms that differ only in the way block columns are orthogonalized and compare their performance with parallel streamed calls to the cuSOLVER 8 \cite{nvidia-cusolver} library routines.

\subsection{Gram Matrix Block Jacobi SVD}
The block Jacobi algorithm is very similar to the vector Algorithm \ref{alg:jacobi}, orthogonalizing pairs of blocks columns instead of vectors. The first method of orthogonalizing pairs of block columns is based on the SVD of their Gram matrix. During the $p$-th sweep, each pair of $m \times k$ block columns $A^{(p)}_i$ and $A^{(p)}_j$ is orthogonalized by forming a $2k \times 2k$ Gram matrix $G^{(p)}_{ij} = {[A^{(p)}_i A^{(p)}_j]}^T [A^{(p)}_i A^{(p)}_j] = {A^{(p)}_{ij}}^T A^{(p)}_{ij}$ and generating a block rotation matrix $U^{(p)}_{ij}$, computed as the left singular vectors of $G^{(p)}_{ij}$ (or equivalently its eigenvectors, since it is symmetric positive definite). Updating $A^{p+1}_{ij} = A^p_{ij} U^{(p)}_{ij}$ orthogonalizes the block columns, since we have
\begin{equation*}
{A^{p+1}_{ij}}^T A^{p+1}_{ij} = {U^{(p)}_{ij}}^T {A^p_{ij}}^T A^p_{ij} U^{(p)}_{ij} = {U^{(p)}_{ij}}^T G^{(p)}_{ij} U^{(p)}_{ij} = \Lambda^{p}_{ij},
\end{equation*}
where $\Lambda^{p}_{ij}$ is a diagonal matrix of the singular values of $G^{(p)}_{ij}$. Orthogonalizing all pairs of block columns until the entire matrix is orthogonal will give us the left singular vectors as the normalized columns and the singular values as the corresponding column norms. If the right singular vectors are needed, we can accumulate the action of the block rotation matrices on the identity matrix. For our batched implementation, we use highly optimized batched \texttt{syrk} and \texttt{gemm} routines from MAGMA to compute $G$ and to apply the block rotations, while the SVD is computed by our shared memory batched kernel. Since different matrices will converge in different numbers of sweeps, we keep track of the convergence of each operation $l$ by computing the norm $e_l$ of the off-diagonal entries of $G$ scaled by its diagonal entries. While this term is an inexact approximation of the off-diagonal terms of the full matrix in each sweep, it is still a good indication of convergence and will cost us at most an extra cheap sweep, since the final sweep will not actually perform any rotations within the SVD of $G$. The entire batched operation will then converge when $e = \max e_l < \epsilon$, where $\epsilon$ is our convergence tolerance. This gives us the Gram matrix path of the batched block Jacobi Algorithm \ref{alg:block_jacobi} to compute the SVD of a batch of matrices in global memory. It is worth noting that the computation of the Gram matrix can be optimized by taking advantage of the special structure of $G$, but since the bulk of the computation is in the SVD of G, it will not result in any significant performance gains. 

\subsection{Direct Block Jacobi SVD}
The Gram matrix method is an indirect way of orthogonalizing block columns and may fail to converge if the matrix is very ill-conditioned. Ill-conditioned matrices can be handled by directly orthogonalizing the columns using their SVD. Since the block columns are rectangular, we first compute their QR decomposition followed by the SVD of the triangular factor $R$. Overwriting the block column $A^p_{ij}$ by the orthogonal factor $Q$ and multiplying it by the left singular vectors of $R$ scaled by the singular values will give us the new block column $A^{p+1}_{ij}$:
\begin{equation*}
A^p_{ij} = Q^p_{ij} R^p_{ij} = \left( Q^p_{ij} U^p_{ij} \Sigma^p_{ij} \right) {V^p_{ij}}^T = A^{p+1}_{ij} {V^p_{ij}}^T.
\end{equation*}
If the right singular vectors are needed, we can accumulate the action of $V^p_{ij}$ on the identity matrix. For our batched implementation, we use the batch QR routine developed in Section \ref{sec:batch_qr} and \texttt{gemm} routines from MAGMA to multiply the orthogonal factor by the left singular vectors, while the SVD is computed by our shared memory batched kernel. The same convergence test used in the Gram matrix method can be used on the triangular factor, since the triangular factor should be close to a diagonal matrix if a pair of block columns are orthogonal. This gives us the direct path of the batched block Jacobi Algorithm \ref{alg:block_jacobi} to compute the SVD of a batch of matrices in global memory.

\begin{algorithm} [t]
\caption{Batched One-sided block Jacobi SVD}
\label{alg:block_jacobi}
	\begin{algorithmic}[1]
		\While {$e > \epsilon$}
			\State $e_l = 0$
			\ForEach {pair of block columns $A_{ij}=[A_i, A_j]$}
				\If{method = GRAM}
					\State $G = \text{batchSyrk}(A_{ij})$
				\Else
					\State $[A_{ij}, G] = \text{batchQR}(A_{ij})$
				\EndIf
				\State $e_l = \text{max}(e_l, \text{scaledOffdiag}(G))$
				\State $U = \text{batchSvd}(G)$
				\State $A_{ij} = \text{batchGemm}(A_{ij}, U)$
			\EndFor
			\State $e = \text{max}(e_l)$
		\EndWhile
	\end{algorithmic}
\end{algorithm}	

\subsection{Performance}
\begin{figure}[h]
	\begin{subfigure}[t]{0.45\textwidth}
	\includegraphics{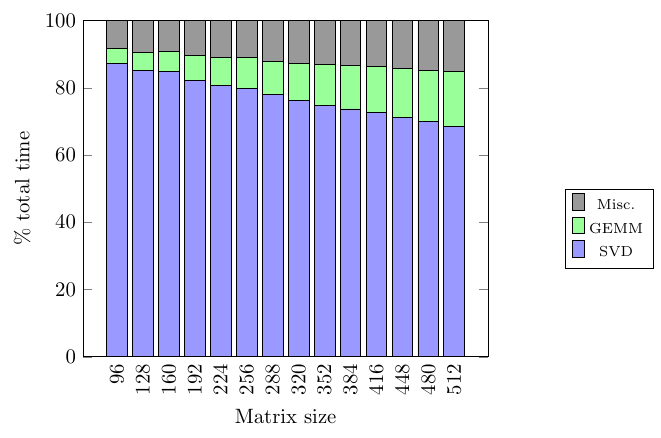}
	\caption{Gram Matrix batched block Jacobi SVD profile.}
	\label{fig:block_jacobi_profile1}
	\end{subfigure}
	\hfill
	\begin{subfigure}[t]{0.45\textwidth}
	\centering
	\includegraphics{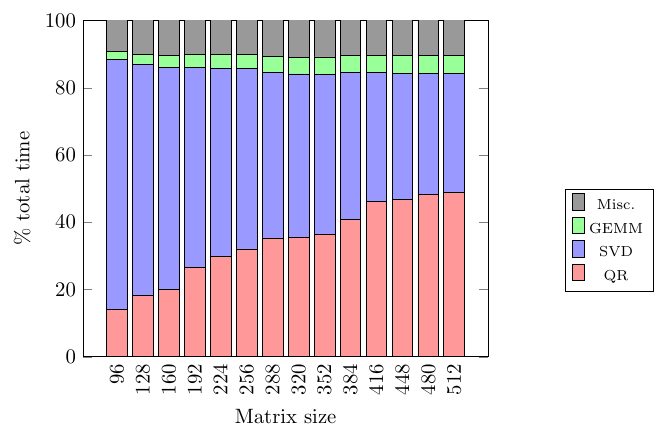}
	\caption{Direct batched block Jacobi SVD profile.}
	\label{fig:block_jacobi_profile2}
	\end{subfigure}
	\caption{Profile of the different phases of the block Jacobi SVD for 200 matrices of varying size on a P100 GPU in double precision. Single precision exhibits similar behavior.}
\label{fig:block_jacobi_profile}
\end{figure}

\begin{figure}
	\begin{subfigure}[t]{0.45\textwidth}
	\includegraphics{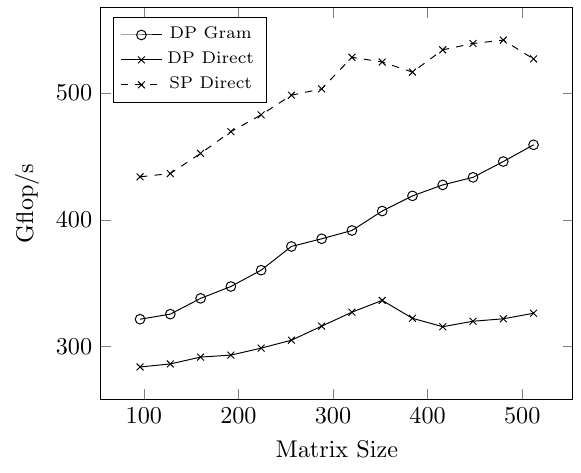}
	\caption{Batched block Jacobi SVD performance.}
	\label{fig:block_jacobi_perf}
	\end{subfigure}
	\hfill
	\begin{subfigure}[t]{0.45\textwidth}
	\centering
	\includegraphics{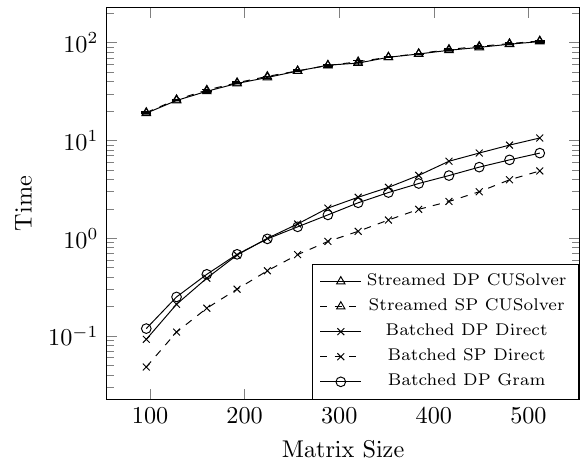}
	\caption{Comparison between streamed cuSOLVER and the batched block Jacobi.}
	\label{fig:osbjvscustream}
	\end{subfigure}
	\caption{Batched block Jacobi performance for 200 matrices of varying size on a P100 GPU in single and double precision arithmetics.}
\end{figure}

Figures \ref{fig:block_jacobi_profile1} and \ref{fig:block_jacobi_profile1} show the profiling of the different computational kernels involved in the batched block algorithms with a block width of $32$, specifically percentages of total execution time for determining convergence and memory operations, matrix multiplications, QR decompositions and the SVD of the Gram matrix. For the Gram matrix approach, the SVD is the most costly phase, even for the larger operations, while the QR and SVD decompositions take almost the same time for the larger matrices in the direct approach. Figure \ref{fig:block_jacobi_perf} shows the performance of the batched block Jacobi SVD of 200 matrices using both methods and Figure \ref{fig:osbjvscustream} compares the performance of our batched SVD routine with a batched routine that uses the cuSOLVER SVD routine using 20 concurrent streams on a P100 GPU. Increasing the number of streams for cuSOLVER showed little to no performance benefits, highlighting the performance limitations of routines that are bound by kernel launch overhead. The matrices are generated randomly using the \texttt{latms} LAPACK routine with a condition number of $10^7$. The Gram matrix approach fails to converge in single precision for these types of matrices, whereas the direct approach always converges; however the Gram matrix approach performs better when it is applicable for the larger matrices due to the strong performance of the matrix-matrix multiplcations. The performance of the block algorithm can be improved by preprocessing the matrix using QR and LQ decompositions to decrease the number of sweeps required for convergence \cite{Oksa_2006} as well as by adaptively selecting pairs of block columns based on the computed offdiagonal norms of their Gram matrices. These changes are beyond the scope of this paper and will be the focus of future work. 

\section{Randomized SVD}
\label{sec:randomized}

As mentioned in Section \ref{sec:intro}, we are often interested in a rank-$k$ approximation of a matrix $A \approx \tilde{U} \tilde{S} \tilde{V}$. We can compute this approximation by first determining the singular value decomposition of the full $m \times n$ matrix $A$ and then truncating the $n-k$ smallest singular values with their corresponding singular vectors; however, when the matrix has low numerical rank $k$, we can obtain the approximation using very fast randomization methods \cite{halko2011finding}. This section will discuss some details of the algorithm and compare its performance with the full SVD using our one-sided block Jacobi kernel. 

\subsection{Implementation}
When the singular values of a matrix decay rapidly, we can compute an approximate SVD using a simple two phase randomization method:
\begin{enumerate}
\item The first phase determines an approximate orthogonal basis $Q$ for the columns of $A$ ensuring that $A \approx QQ^T A$. When the numerical rank $k$ of $A$ is low, we can be sure that $Q$ has a small number of columns as well. In \cite{halko2011finding} we see that by drawing $k+p$ sample vectors $y = Aw$ from random input vectors $w$, we can obtain a reliable approximate basis for $A$ which can then be orthogonalized. This boils down to computing a matrix $Y = A \Omega$, where $\Omega$ is a $n \times (k + p)$ random Gaussian sampling matrix, and then computing the QR decomposition of $Y = Q R_y$, where $Q$ is the desired approximate orthogonal basis. 
\item The second phase uses the fact that $A \approx QQ^T A$ to compute a matrix $B = Q^T A$ so that we now have $A \approx QB$. Forming the SVD of $B = U_B S V^T$, we finalize our approximation $A \approx QU_B S V^T = U S V^T$. For the wide $(k+p) \times n$ matrix $B$, we can first compute a QR decomposition of its transpose, followed by the SVD of the upper triangular factor. 
\end{enumerate}
Algorithm \ref{alg:batch_rsvd} shows that the core computations for the randomized method are matrix-matrix multiplications, QR decompositions, and the singular value decompositions of small matrices. Using the batched routines from the previous sections, it is straightforward to form the required randomized batched SVD. More robust randomized SVD algorithms would employ randomized subspace iteration methods to obtain a better basis $Q$ for the columns of $A$ and rely on these same core kernels, but will not be further discussed here. 

\begin{algorithm} [t]
\caption{Batched Randomized SVD}
\label{alg:batch_rsvd}
	\begin{algorithmic}[1]
		\Procedure{RSVD}{$A, k, p$}
			\State $[m, n] = size(A)$
			\State $\Omega = \text{Rand}(n, k + p)$
			\State $Y = \text{batchGemm}(A, \Omega)$
			\State $[Q, R_y] = \text{batchQR}(Y)$
			\State $B = \text{batchGemm}(Q^T, A)$
			\State $[Q_B, R_B] = \text{batchQR}(B^T)$
			\State $[U_R, S, V_R] = \text{batchSvd}(R_B^T)$
			\State $U = \text{batchGemm}(Q, U_R)$
			\State $V = \text{batchGemm}(Q_B, V_R)$
		\EndProcedure
	\end{algorithmic}
\end{algorithm}	

\subsection{Performance}
\begin{figure}[h]
	\centering
	\includegraphics{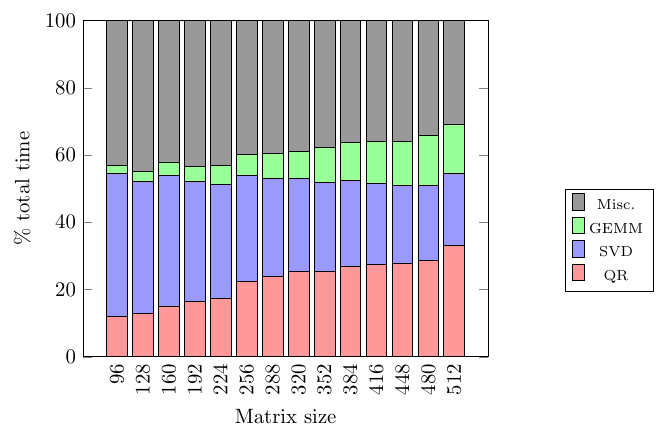}
\caption{Profile of the different phases of the batched randomized SVD for 200 matrices of varying size on a P100 GPU in double precision. Single precision exhibits similar behavior.}
\label{fig:rsvd_profile}
\end{figure}

\begin{figure}
	\begin{subfigure}[t]{0.45\textwidth}
	\centering
	\includegraphics{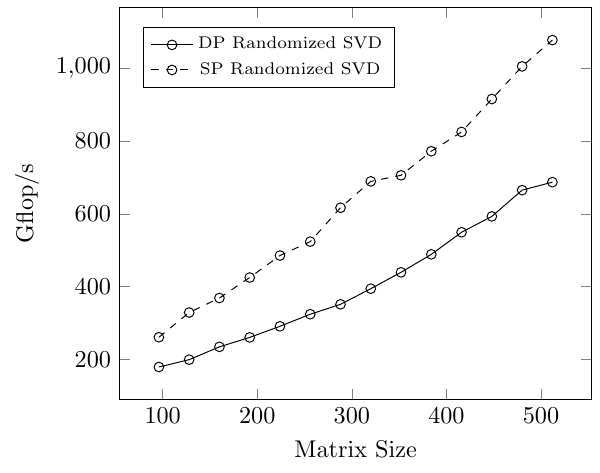}
	\caption{Batched randomized SVD performance.}
	\label{fig:rsvd_perf}
	\end{subfigure}
	\hfill
	\begin{subfigure}[t]{0.45\textwidth}
	\centering	
	\includegraphics{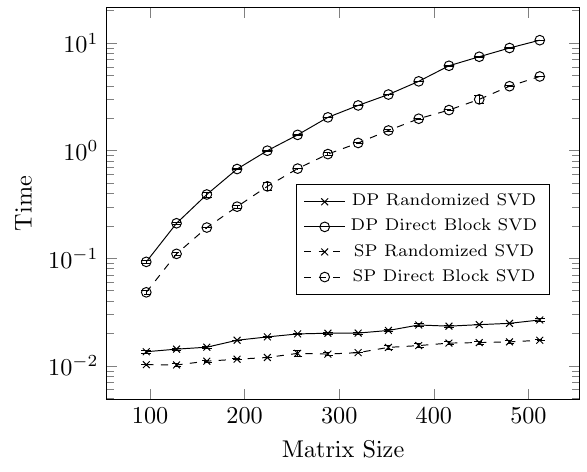}
	\caption{Comparison between the batched block Jacobi and the batched randomized SVD.}
	\label{fig:rsvd_vs_osbj}
	\end{subfigure}
	\caption{Batched randomized SVD performance for 200 matrices of varying size on a P100 GPU in single and double precision for the first 64 singular values and vectors.}
\end{figure}

Figure \ref{fig:rsvd_profile} shows the profiling of the different kernels used in the randomized batched routine for determining the top 64 singular values and vectors of randomly generated low rank matrices using the \texttt{latms} LAPACK routine. The miscellaneous portion includes random number generation using the CURAND library's default random number generator and a Gaussian distribution, batched transpose operations and memory operations. We can see that the performance of all kernels play almost equally important roles in the performance of the randomized routine as the matrix size grows while keeping the computed rank the same. Figure \ref{fig:rsvd_perf} shows the performance of the batched randomized SVD of 200 operations and Figure \ref{fig:rsvd_vs_osbj} compares the runtimes of the direct block one-sided Jacobi routine with the randomized SVD on a P100 GPU for the same set of matrices, showing that significant time savings can be achieved even for relatively small blocks. 

\section{Application to Hierarchical Matrix Compression}
\label{sec:application}

As an application of the batched kernels presented, we consider the problem of compressing/recompressing hierarchical matrices. This is a problem of significant importance for building hierarchical matrix algorithms and in fact was our primary motivation for the development of the batched kernels.

Hierarchical matrices~\cite{hackbush_2000, hackbush_h2_2000, hackbush_1999} have received substantial attention in recent years because of their ability to store and perform algebraic operations in near linear complexity rather than the $O(n^2)$ and $O(n^3)$ that regular dense matrices require. The effectiveness of hierarchical matrices comes from the fact they can approximate a matrix by a (quad)-tree of blocks where many of the blocks in the off-diagonal regions have a rapidly decaying spectrum and can therefore be well-approximated by numerically low rank representations. It is these low rank representations, at different levels of the hierarchical tree, that reduce the memory footprint and operations complexity of the associated matrix algorithms. Hackbush~\cite{hbook} shows that many of the large dense matrices that appear in scientific computing, such as from the discretization of integral operators, Schur complements of discretized PDE operators, and covariance matrices, can be well approximated by these hierarchical representations. 

Reviewing and analyzing hierarchical matrix algorithms is beyond the scope of this paper. Here we focus on the narrow task of compressing hierarchical matrices. This compression task may be viewed as a generalization of the well-known compression (i.e., low rank approximation) of large dense matrices to the case of hierarchical matrices. For large dense matrices, one way to perform the compression is to generate a single exact or approximate SVD ($U \Sigma V^T$) and truncate the spectrum $\Sigma$ to the desired tolerance, to produce a truncated or ``compressed'' representation $(\bar{U} \bar{\Sigma} \bar{V}^T)$. For hierarchical matrices, the equivalent operations involve \emph{batched SVDs} on small blocks, with one batched kernel call per level of the tree in the hierarchical representation. The size of the batch in every such call is the number of nodes at the corresponding level in the tree. 

Compression algorithms with controllable accuracy are important practically, because it is often the case that the hierarchical matrices generated by analytical methods can be compressed with no significant loss of accuracy. Even more importantly, when performing matrix operations such as additiona and multiplication, the apparent ranks of the blocks often grow and have to be recompressed regularly during the operations to prevent superlinear growth in memory requirements. 

\begin{figure}
\begin{subfigure}[t] {0.45\linewidth}
	\centering
    \includegraphics[width=0.9\linewidth]{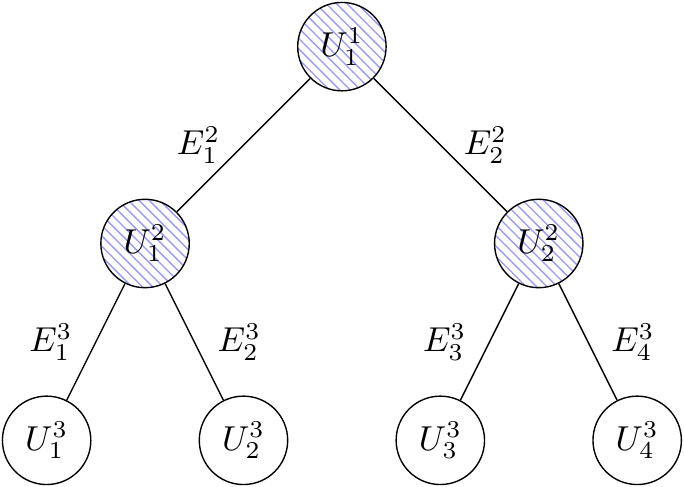}
    \caption{Basis tree $U$ of an $\mathcal{H}^2$-matrix. Leaf nodes are stored explicitly, whereas inner nodes are represented implicitly using the transfer matrices $E$.}
    \label{fig:basis_tree}
\end{subfigure}
\hfill
\begin{subfigure}[t] {0.45\linewidth}
	\centering
    \includegraphics[width=0.8\linewidth]{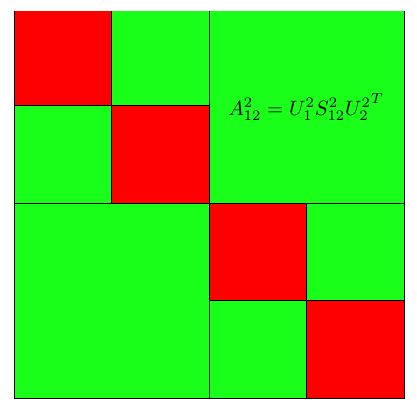}
    \caption{Leaves of matrix tree for a simple hierarchical matrix. Red blocks represent dense leaves and green blocks are low rank leaves.}
	\label{fig:hmatrix}
\end{subfigure}
\caption{The basis tree and matrix tree leaves of a simple $\mathcal{H}^2$-matrix.}
\end{figure}

\subsection{$\mathcal{H}^2$-matrix representation}

For our application, we use the memory efficient $\mathcal{H}^2$ variant of hierarchical matrices which exhibit linear complexity in time and space for many of its core operations. In the $\mathcal{H}^2$-matrix format, a hierarchical matrix is actually represented by three trees: 
\begin{itemize}
\item Row and column basis column trees $U$ and $V$ that organize the row and column indices of the matrix hierarchically. Each node represents a set of basis vectors for the row and column spaces of the blocks of $A$. Nodes at the leaves of the tree store these vectors explicitly, while inner nodes store only transfer matrices that allow us to implicitly represent basis vectors in terms of their children. A basis tree with this parent-child relationship of the nodes is called a nested basis. For example, in a binary row basis tree $U$ with transfer matrices $E$, we can explicitly compute the basis vectors for a node $i$ with children $i_1$ and $i_2$ at level $l$ as:
\begin{equation*} 
	U^{l-1}_i = 
	\begin{bmatrix} U^l_{i_1} & \\ & U^l_{i_2} \end{bmatrix}
	\begin{bmatrix} E^l_{i_1} \\ E^l_{i_2} \end{bmatrix}. 
\end{equation*} 
Figure \ref{fig:basis_tree} shows an example of a binary basis tree.

\item A matrix tree for the hierarchical blocking of $A$ formed by a dual traversal of the nodes of the two basis trees. A leaf is determined when the block is either small enough and stored as an $m \times m$ dense matrix, or when a low rank approximation of the block meets a specified accuracy tolerance. For the latter case, the node is stored as a $k_l \times k_l$ coupling matrix $S$ at each level $l$ of the tree, where $k_l$ is the rank at level $l$. The block $A_{ts}$ of the matrix, where $t$ is the index set of a node in the row basis tree $U$ and $s$ is the index set of a node in the column basis $V$, is then approximated as $A_{ts} \approx U_t S_{ts} V_s^T$. Figure \ref{fig:hmatrix} shows the leaves of the matrix quadtree of a simple hierarchical matrix. 
\end{itemize}
For the case of symmetric matrices, the $U$ and $V$ trees are identical. Our numerical results below are from a symmetric covariance matrix. 

\subsection{Compression}
The compression of a symmetric $\mathcal{H}^2$-matrix $A_H$, represented by the two trees $U$ (with its transfer matrices $E$) and $S$, involves generating a new optimal basis tree $\widetilde{U}$ (with its transfer matrices $\widetilde{E}$) in a truncation phase, and a new $\widetilde{S}$ that expresses the contents of the matrix blocks in this new basis in a projection phase. 

We present a version of the truncation algorithm that generates a memory efficient basis $[\widetilde{U}, \widetilde{E}]$ from a representation of the matrix in a given $[U, E]$ basis. More sophisticated algebraic compression algorithms that involve the use of $S$ in the truncation phase in order to generate a more efficient basis will be the subject of future work. 

The truncation phase computes the SVD of the nodes of the basis tree $U$ level by level, with all nodes in a level being processed in parallel to produce the new basis $\widetilde{U}$. We have an explicit representation of the basis vectors at the leaves, so we can compute the SVD of all leaf nodes in parallel with our batched kernels and truncate the singular vectors whose singular values are lower than our relative compression threshold $\epsilon$. Truncating the node to the relative threshold using the SVD will give us an approximation of the leaf such that $\frac{||U - \widetilde{U}||_F}{||U||_F} \le \epsilon$. With the new leaf nodes, we can compute projection matrices in a tree $T$, where each node $i$, $T^d_i = \widetilde{U^d_i}^T U^d_i$ and $d$ is the leaf level. Sweeping up the tree, we process the inner nodes while preserving the nested basis property. Using the parent-child relationship of a node $i$ with children $i_1$ and $i_2$ at level $l$, we have:
\begin{equation*}
U^{l-1}_i = \begin{bmatrix}
U^l_{i_1}  & 		\\
& U^l_{i_2}
\end{bmatrix}\begin{bmatrix}
E^l_{i_1}\\
E^l_{i_2}
\end{bmatrix} \approx
\begin{bmatrix}
\widetilde{U}^l_{i_1}  & 		\\
& \widetilde{U}^l_{i_2}
\end{bmatrix}\begin{bmatrix}
T^l_{i_1} E^l_{i_1}\\
T^l_{i_2} E^l_{i_2}
\end{bmatrix} = 
\begin{bmatrix}
\widetilde{U}^l_{i_1}  & 		\\
& \widetilde{U}^l_{i_2}
\end{bmatrix} TE_i
\end{equation*}
After forming the $TE$ matrices using batched matrix-matrix multiplication, we compute their SVD $TE = QSW^T$ using the batched SVD kernel and truncate as we did for the leaves to form the truncated $\widetilde{TE}$ matrices as:
\begin{equation*}
\widetilde{TE}_i = \widetilde{Q}_i \left( \widetilde{S}_i \widetilde{W}_i^T \right) = \begin{bmatrix}
\widetilde{E}^l_{i_1}\\
\widetilde{E}^l_{i_2}
\end{bmatrix} T^{l-1}_i
\end{equation*}
where $\widetilde{E}^l$, the block rows of $\widetilde{Q}$, are the new transfer matrices at level $l$ of our compressed nested basis and $T^{l-1}$ are the projection matrices for level $(l-1)$. The key computations  involved in this truncation phase consist then of one batched SVD involving the leaves of the tree, followed by a sequence of batched SVDs, one per level of the tree, involving the transfer matrices and data from the lower levels. 

The projection phase consists of transforming the coupling matrices in the matrix tree using the generated projection matrices of the truncation phase. For each coupling matrix $S_{ts}$, we compute a new coupling matrix $\widetilde{S}_{ts} = T_t S_{ts} T_s^T$ using batched matrix-matrix multiplications. This phase of the operation consumes much less time than the truncation phase on GPUs, because of substantial efficiencies in executing regular arithmetically intensive operations on them. 

\subsection{Results}
\begin{figure}[h]
	\begin{subfigure}[t]{0.45\textwidth}
	\centering	
	\includegraphics{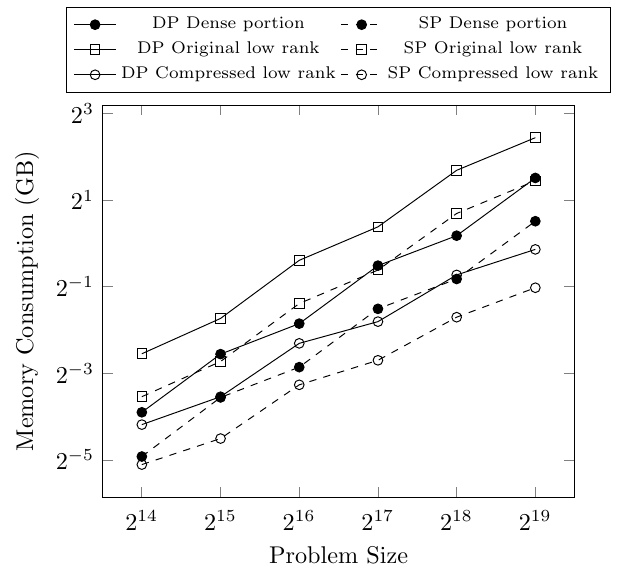}
	\caption{Memory savings.}
	\label{fig:compression}
	\end{subfigure}
	\hfill
	\begin{subfigure}[t]{0.45\textwidth}
	\centering	
	\includegraphics{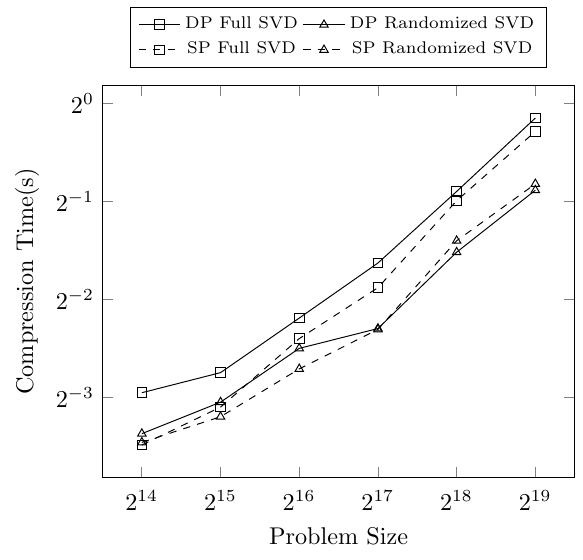}
	\caption{Compression time using randomized SVD with 32 samples and the full SVD using the shared memory kernel.}
	\label{fig:compression_time}
	\end{subfigure}
	\caption{Compression results for sample covariance matrices generated from 2D spatial statistics on a P100 GPU in single and double precision, using a relative Frobenius norm threshold of $10^{-7}$ and initial rank 64.}
\end{figure}

\begin{figure}[h]
	\centering	
	\includegraphics{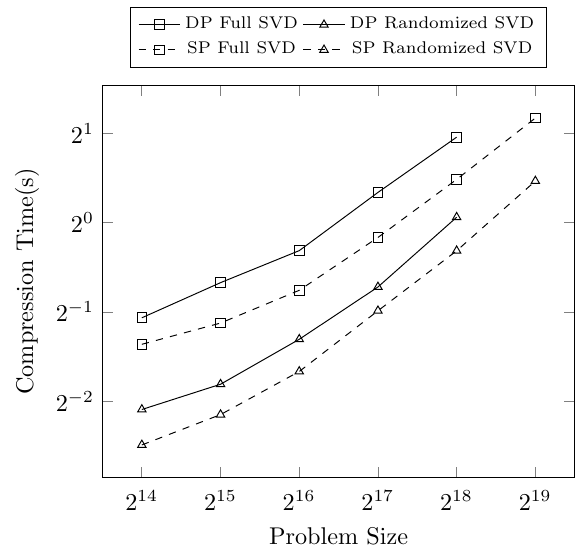}
	\caption{Compression time for a coarser matrix tree with initial rank 121 comparing the randomized SVD with 64 samples and the full SVD.}
	\label{fig:compression_time2}
\end{figure}

As an illustration of the effectiveness of the algebraic compression procedure, we generate covariance matrices of various sizes for a spatial Gaussian process with $n$ observation points placed on a random perturbation of a regular discretization of the unit square $[0,1] \times [0,1]$ and an isotropic exponential kernel with correlation length of $0.1$. Hierarchical representations of the formally dense $n \times n$ covariance matrices are formed analytically by first clustering the points in a KD-tree using a mean split giving us the hierarchical index sets of the basis tree. The basis vectors and transfer nodes are generated using Chebyshev interpolation \cite{borm2007approximating}. The matrix tree is constructed using a dual traversal of the basis tree \cite{hackbush_2000, hackbush_2003}, and the coupling matrices are generated by evaluating the kernel at the interpolation points. The approximation error of the constructed matrix is then controlled by varying the number of interpolation points and by varying the leaf admissibility condition during the dual tree traversal. An approximation error of $10^{-7}$ has been used in the following tests and a relative truncation error $\epsilon=\frac{||A_H - \widetilde{A}_H||_F}{||A_H||_F} \le 10^{-7}$ has been used to maintain the accuracy of the compressed matrices. Figure \ref{fig:compression} shows the memory consumption before and after compression of hierachical covariance matrices with leaf size $64$ and initial rank $64$ (corresponding to an $8 \times 8$ Chebyshev grid). The dense part remains untouched, while the low rank part of the representation sees a substantial decrease in memory consumption after compression with minimal loss of accuracy. Figure \ref{fig:compression_time} shows the expected asymptotic linear growth in time of the compression algorithm and shows the effect of using the randomized SVD with $32$ samples instead of the full SVD as computed by the shared memory kernel. Figure \ref{fig:compression_time2} shows another example where the admissibility condition is weakened to generate a coarser matrix tree with an increased rank of 121 (corresponding to an $11 \times 11$ Chebyshev grid) and the randomized SVD with $64$ samples also reduces compression time when compared to the full SVD using the direct block Jacobi kernels. 

\section{Conclusions and Future Work}
\label{sec:conclusion}
In this paper, we described the implementation of efficient batched kernels for the QR decomposition and randomized singular value decomposition of low rank matrices hosted on the GPU. Our batched QR kernel provides significant performance improvements for small matrices over existing state of the art libraries, and our batched SVD routines are the first of their kind on the GPU, with performance exceeding 800/400 GFLOP/s on a batch of 1000 matrices of size $512 \times 512$ in single/double precision. We illustrated the power of these kernels on a problem involving the algebraic compression of hierarchical matrices stored entirely in GPU memory, and demonstrated a high-performance compression algorithm yielding significant memory savings on practical problems. In the future, we plan to investigate alternatives to the one-sided Jacobi algorithm for the SVD of the small blocks in the randomized algorithm and improve the performance of the blocked algorithms using preconditioning and adaptive block column pair selection. We also plan to develop a suite of hierarchical matrix operations suited for execution on modern GPU and manycore architectures.

\section*{Acknowledgments}
We thank the NVIDIA Corporation for providing access to the P100 GPU used in this work.

\bibliography{arxiv_batch_svd}

\begin{thebibliography}{10}
\providecommand{\url}[1]{#1}
\csname url@samestyle\endcsname
\providecommand{\newblock}{\relax}
\providecommand{\bibinfo}[2]{#2}
\providecommand{\BIBentrySTDinterwordspacing}{\spaceskip=0pt\relax}
\providecommand{\BIBentryALTinterwordstretchfactor}{4}
\providecommand{\BIBentryALTinterwordspacing}{\spaceskip=\fontdimen2\font plus
\BIBentryALTinterwordstretchfactor\fontdimen3\font minus
  \fontdimen4\font\relax}
\providecommand{\BIBforeignlanguage}[2]{{%
\expandafter\ifx\csname l@#1\endcsname\relax
\typeout{** WARNING: IEEEtran.bst: No hyphenation pattern has been}%
\typeout{** loaded for the language `#1'. Using the pattern for}%
\typeout{** the default language instead.}%
\else
\language=\csname l@#1\endcsname
\fi
#2}}
\providecommand{\BIBdecl}{\relax}
\BIBdecl

\bibitem{halko2011finding}
N.~Halko, P.-G. Martinsson, and J.~A. Tropp, ``Finding structure with
  randomness: Probabilistic algorithms for constructing approximate matrix
  decompositions,'' \emph{SIAM Review}, vol.~53, no.~2, pp. 217--288, 2011.

\bibitem{golub2013matrix}
G.~Golub and C.~Van~Loan, \emph{Matrix Computations}.\hskip 1em plus 0.5em
  minus 0.4em\relax Johns Hopkins University Press, 2013.

\bibitem{trefethen1997numerical}
L.~Trefethen and D.~Bau, \emph{Numerical Linear Algebra}.\hskip 1em plus 0.5em
  minus 0.4em\relax Society for Industrial and Applied Mathematics, 1997.

\bibitem{demmel1992jacobi}
J.~Demmel and K.~Veselic, ``Jacobi's method is more accurate than {QR},''
  \emph{SIAM Journal on Matrix Analysis and Applications}, vol.~13, no.~4, pp.
  1204--1245, 1992.

\bibitem{batchqr_haidar}
A.~Haidar, T.~T. Dong, S.~Tomov, P.~Luszczek, and J.~Dongarra, ``A framework
  for batched and {GPU}-resident factorization algorithms applied to block
  {H}ouseholder transformations.'' in \emph{ISC}, ser. Lecture Notes in
  Computer Science, J.~M. Kunkel and T.~Ludwig, Eds., vol. 9137.\hskip 1em plus
  0.5em minus 0.4em\relax Springer, 2015, pp. 31--47.

\bibitem{batch_haidar}
A.~Haidar, T.~Dong, P.~Luszczek, S.~Tomov, and J.~Dongarra, ``Optimization for
  performance and energy for batched matrix computations on {GPU}s,'' in
  \emph{Proceedings of the 8th Workshop on General Purpose Processing Using
  GPUs}, ser. GPGPU-8.\hskip 1em plus 0.5em minus 0.4em\relax New York, NY,
  USA: ACM, 2015, pp. 59--69.

\bibitem{wilt2013cuda}
N.~Wilt, \emph{The {CUDA} Handbook: A Comprehensive Guide to {GPU}
  Programming}.\hskip 1em plus 0.5em minus 0.4em\relax Pearson Education, 2013.

\bibitem{volkov2010better}
V.~Volkov, ``Better performance at lower occupancy,'' \emph{Proceedings of the
  GPU technology conference, GTC}, vol.~10, 2010.

\bibitem{charara_batch_tdla}
\BIBentryALTinterwordspacing
A.~Charara, D.~E. Keyes, and H.~Ltaief, ``{Batched Triangular Dense Linear
  Algebra Kernels for Very Small Matrix Sizes on GPUs},'' \emph{Submitted to
  ACM Transactions on Mathematical Software}. [Online]. Available:
  \url{http://hdl.handle.net/10754/622975}
\BIBentrySTDinterwordspacing

\bibitem{caqr_anderson}
M.~Anderson, G.~Ballard, J.~Demmel, and K.~Keutzer, ``Communication-avoiding
  {QR} decomposition for {GPU}s,'' in \emph{Parallel Distributed Processing
  Symposium {(IPDPS)}, 2011 {IEEE} International}, May 2011, pp. 48--58.

\bibitem{kotas_svd}
C.~Kotas and J.~Barhen, ``Singular value decomposition utilizing parallel
  algorithms on graphical processors,'' in \emph{OCEANS'11 MTS/IEEE KONA}, Sept
  2011, pp. 1--7.

\bibitem{Kang2015}
H.-P. Kang and C.-R. Lee, \emph{Improving Performance of Convolutional Neural
  Networks by Separable Filters on GPU}.\hskip 1em plus 0.5em minus 0.4em\relax
  Berlin, Heidelberg: Springer Berlin Heidelberg, 2015, pp. 638--649.

\bibitem{badolato_2015}
I.~Badolato, L.~d.~Paula, and R.~Farias, ``Many svds on gpu for image mosaic
  assemble,'' in \emph{2015 International Symposium on Computer Architecture
  and High Performance Computing Workshop (SBAC-PADW)}, Oct 2015, pp. 37--42.

\bibitem{tnld10}
S.~Tomov, R.~Nath, H.~Ltaief, and J.~Dongarra, ``Dense linear algebra solvers
  for multicore with {GPU} accelerators,'' in \emph{Proc. of the IEEE
  IPDPS'10}.\hskip 1em plus 0.5em minus 0.4em\relax Atlanta, GA: IEEE Computer
  Society, April 19-23 2010, pp. 1--8, {DOI:~10.1109/IPDPSW.2010.5470941}.

\bibitem{nvidia-cublas}
\BIBentryALTinterwordspacing
{NVIDIA}, \emph{{CUBLAS Library User Guide}}, v8.0~ed.,
  http://docs.nvidia.com/cuda/cublas, {NVIDIA}, 2017. [Online]. Available:
  \url{http://docs.nvidia.com/cuda/cublas}
\BIBentrySTDinterwordspacing

\bibitem{warp_shfl}
J.~Cheng, M.~Grossman, and T.~McKercher, \emph{Professional {CUDA} {C}
  Programming}, ser. EBL-Schweitzer.\hskip 1em plus 0.5em minus 0.4em\relax
  Wiley, 2014.

\bibitem{journals/concurrency/KurzakLDB10}
\BIBentryALTinterwordspacing
J.~Kurzak, H.~Ltaief, J.~Dongarra, and R.~M. Badia, ``Scheduling dense linear
  algebra operations on multicore processors,'' \emph{Concurrency and
  Computation: Practice and Experience}, vol.~22, no.~1, pp. 15--44, 2010.
  [Online]. Available: \url{http://dx.doi.org/10.1002/cpe.1467}
\BIBentrySTDinterwordspacing

\bibitem{parosbj_zhou}
B.~B. Zhou and R.~P. Brent, ``On parallel implementation of the one-sided
  {J}acobi algorithm for singular value decompositions,'' in \emph{Parallel and
  Distributed Processing, 1995. Proceedings. Euromicro Workshop on}, Jan 1995,
  pp. 401--408.

\bibitem{ZHOU19971}
B.~Zhou and R.~Brent, ``A parallel ring ordering algorithm for efficient
  one-sided {J}acobi {SVD} computations,'' \emph{Journal of Parallel and
  Distributed Computing}, vol.~42, no.~1, pp. 1 -- 10, 1997.

\bibitem{bevcka1999block1}
M.~BE{\v{C}}KA and M.~Vajter{\v{s}}ic, ``Block-{J}acobi {SVD} algorithms for
  distributed memory systems {I}: Hypercubes and rings*,'' \emph{Parallel
  Algorithms and Applications}, vol.~13, no.~3, pp. 265--287, 1999.

\bibitem{bevcka1999block2}
------, ``Block-{j}acobi {svd} algorithms for distributed memory systems {II}:
  Meshes∗,'' \emph{Parallel Algorithms and Applications}, vol.~14, no.~1, pp.
  37--56, 1999.

\bibitem{bevcka2015new}
M.~Be{\v{c}}ka, G.~Ok{\v{s}}a, and M.~Vajter{\v{s}}ic, ``New dynamic orderings
  for the parallel one--sided block-{J}acobi {SVD} algorithm,'' \emph{Parallel
  Processing Letters}, vol.~25, no.~02, p. 1550003, 2015.

\bibitem{nvidia-cusolver}
\BIBentryALTinterwordspacing
{NVIDIA}, \emph{{cuSOLVER Library User Guide}}, v8.0~ed.,
  http://docs.nvidia.com/cuda/cusolver/, {NVIDIA}, 2017. [Online]. Available:
  \url{http://docs.nvidia.com/cuda/cusolver}
\BIBentrySTDinterwordspacing

\bibitem{Oksa_2006}
\BIBentryALTinterwordspacing
G.~Ok\v{s}a and M.~Vajter\v{s}ic, ``Efficient pre-processing in the parallel
  block-{J}acobi {SVD} algorithm,'' \emph{Parallel Comput.}, vol.~32, no.~2,
  pp. 166--176, Feb. 2006. [Online]. Available:
  \url{http://dx.doi.org/10.1016/j.parco.2005.06.006}
\BIBentrySTDinterwordspacing

\bibitem{hackbush_2000}
W.~Hackbusch and B.~N. Khoromskij, ``\BIBforeignlanguage{English}{A sparse
  {$\mathcal{H}$}-matrix arithmetic. {P}art {II}: Application to
  multi-dimensional problems},''
  \emph{\BIBforeignlanguage{English}{Computing}}, vol.~64, no.~1, pp. 21--47,
  2000.

\bibitem{hackbush_h2_2000}
W.~Hackbusch, B.~Khoromskij, and S.~Sauter, ``\BIBforeignlanguage{English}{On
  {$\mathcal{H}^2$}-matrices},'' in \emph{\BIBforeignlanguage{English}{Lectures
  on Applied Mathematics}}, H.-J. Bungartz, R.~Hoppe, and C.~Zenger, Eds.\hskip
  1em plus 0.5em minus 0.4em\relax Springer Berlin Heidelberg, 2000, pp. 9--29.

\bibitem{hackbush_1999}
W.~Hackbusch, ``\BIBforeignlanguage{English}{A sparse matrix arithmetic based
  on {$\mathcal{H}$}-matrices. {P}art {I}: Introduction to
  {$\mathcal{H}$}-matrices},'' \emph{\BIBforeignlanguage{English}{Computing}},
  vol.~62, no.~2, pp. 89--108, 1999.

\bibitem{hbook}
------, \emph{{Hierarchical matrices : Algorithms and Analysis}}, ser. Springer
  series in computational mathematics.\hskip 1em plus 0.5em minus 0.4em\relax
  Berlin: Springer, 2015, vol.~49.

\bibitem{borm2007approximating}
S.~B{\"o}rm and J.~Garcke, ``Approximating gaussian processes with
  {$\mathcal{H}^2$}-matrices,'' in \emph{European Conference on Machine
  Learning}.\hskip 1em plus 0.5em minus 0.4em\relax Springer, 2007, pp. 42--53.

\bibitem{hackbush_2003}
L.~Grasedyck and W.~Hackbusch, ``\BIBforeignlanguage{English}{Construction and
  arithmetics of {$\mathcal{H}$}-matrices},''
  \emph{\BIBforeignlanguage{English}{Computing}}, vol.~70, no.~4, pp. 295--334,
  2003.

\end{thebibliography}
\bibliographystyle{ieeetran}

\end{document}